\documentclass{PoS}
\usepackage{amssymb,amsmath,bm,bbm}
\usepackage{epsf,epsfig}
\usepackage{afterpage}
\usepackage{longtable}
\usepackage{cite}
\usepackage{latexsym, mathrsfs}
\usepackage{graphics}
\usepackage{url}
\usepackage{paralist}
\usepackage{bbold}

\newcommand{\be}{\begin{equation}}
\newcommand{\ee}{\end{equation}}
\newcommand{\bea}{\begin{eqnarray}}
\newcommand{\eea}{\end{eqnarray}}
\newcommand{\bec}{\begin{center}}
\newcommand{\eec}{\end{center}}

\title{Investigating thermalization of a strongly interacting non-Abelian plasma}

\ShortTitle{Investigating thermalization of a strongly interacting non-Abelian plasma}

\author{Loredana Bellantuono\\
   Institute of Physics, Jagiellonian University, Lojasiewicza 11, 30-348 Krakow, Poland\\
       E-mail: \email{loredana.bellantuono@ba.infn.it}}
       
       \author{Pietro Colangelo\\
 Istituto Nazionale di Fisica Nucleare - Sezione di Bari,  via Orabona 4, 70125 Bari, Italy\\
       E-mail: \email{pietro.colangelo@ba.infn.it}}
       
       \author{\speaker{Fulvia De Fazio}\\
        Istituto Nazionale di Fisica Nucleare - Sezione di Bari,  Italy\\
        E-mail: \email{fulvia.defazio@ba.infn.it}}

\author{Floriana Giannuzzi\\
Istituto Nazionale di Fisica Nucleare - Sezione di Bari,  Italy\\
       E-mail: \email{floriana.giannuzzi@ba.infn.it}}
       
\author{Stefano Nicotri\\
 Istituto Nazionale di Fisica Nucleare - Sezione di Bari, Italy\\
        E-mail: \email{stefano.nicotri@ba.infn.it}}

\abstract{Using gauge/gravity duality  methods, we  study the relaxation towards equilibrium of  strongly interacting non-Abelian matter. We adopt  boundary sourcing to drive the system out-of-equilibrium, and  analyze  the equilibration process through  local probes (energy density and pressures) and nonlocal probes (lengths of the geodesics between two boundary points and  extremal surfaces having a Wilson loop as contour on the boundary).
We also  investigate  the real-time dissociation of the heavy quarkonium in the out-of-equilibrium plasma. Systematic comparison with the results of a geometry dual to viscous hydrodynamics sheds light on the thermalization process.
}

\FullConference{EPS-HEP 2017, European Physical Society conference on High Energy Physics\\
		5-12 July 2017\\
		Venice, Italy}

\begin{document}

\section{Introduction}
Within an elapsed  time  of ${\cal O}(1\, {\rm fm})$ the strongly interacting matter produced in relativistic heavy ion  collisions evolves from the far-from-equilibrium state to a  viscous hydrodynamic regime \cite{Braun-Munzinger:2015hba}.  The description of the pre-equilibrium configuration and of the transition to the hydrodynamic behavior  is an issue difficult to face theoretically \cite{Ruggieri:2017ioa}. Suitable methods
make use of the gauge/gravity duality approach, invoking a correspondence between  a strongly coupled conformal field theory  on a 4D Minkowski space ${\cal M}_4$, and a weakly
 coupled gravity theory on AdS$_5 \times $S$_5$, the {\it bulk},  of which ${\cal M}_4$ is the boundary  \cite{Maldacena:1997re,Witten:1998qj,Gubser:1998bc}. Thermalization in the boundary  theory 
 corresponds to the formation of a   black-hole  in the bulk with time-dependent horizon, and it
 can be probed considering the  boundary theory stress-energy tensor $T_{\mu \nu}$, as well as correlation functions of boundary theory operators.
 
Among the 4D coordinates $x^\mu=(x^0,\,x^1,\,x^2,\,x^3)$,  we identify $x^3=x_\parallel$ with the collision axis. Writing   $x^0=\tau \, \cosh y$ and $x^3=\tau \, \sinh y$ in terms of proper time $\tau$ and  rapidity $y$, the  Minkowski metric reads:
$
ds^2=-d\tau^2+\tau^2 dy^2+dx_\perp^2$, with $x_\perp =\{x^1,\,x^2\}$.
Boost-invariance along $x_\parallel$, rotation and translation  invariance in the $x_\perp$  plane imply for an expanding inviscid fluid that  $T_{\mu \nu}$ is diagonal, with components (energy density, transverse and longitudinal pressures) depending only on $\tau$:
$\epsilon(\tau) = \displaystyle\frac{c}{\tau^{4/3}}$ and $
p_\parallel = -\epsilon(\tau) -\tau \epsilon^\prime (\tau)$, 
$p_\perp = \epsilon(\tau) +\tau  \epsilon^\prime (\tau)/2$,
with $c$  a constant  \cite{Bjorken:1982qr}.
Corrections for  viscous hydrodynamics  modify these relations:
\bea
\epsilon(\tau)&=& \frac{3 \pi^4 \Lambda^4}{4 (\Lambda \tau)^{4/3} }\left[ 1-\frac{2c_1}{ (\Lambda \tau)^{2/3}}+\frac{c_2}{ (\Lambda \tau)^{4/3}} + {\cal O}\left( \frac{1}{(\Lambda \tau)^2} \right )\right] \label{hydroeps} \\
p_\parallel (\tau)&=&\frac{ \pi^4 \Lambda^4}{ 4(\Lambda \tau)^{4/3} } \left[ 1-\frac{6c_1}{ (\Lambda \tau)^{2/3}}+\frac{5c_2}{ (\Lambda \tau)^{4/3}} + {\cal O}\left( \frac{1}{(\Lambda \tau)^2} \right )\right] \label{hydroppar} \\
p_\perp (\tau) &=& \frac{ \pi^4 \Lambda^4}{ 4(\Lambda \tau)^{4/3} } \left[ 1-\frac{c_2}{ (\Lambda \tau)^{4/3}} + {\cal O}\left( \frac{1}{(\Lambda \tau)^2} \right ) \right] \,\,\, ,\label{hydropperp}
\eea
with $c_{1,2}$ numerical constants and $\Lambda$  a parameter   \cite{Janik:2005zt}.
An  effective fluid temperature can be defined:  $\epsilon (\tau)=\frac{3}{4} \pi^4 T_{eff}(\tau)^4$.
Invoking  the gauge/gravity correspondence, the dual of  $T_{\mu \nu}$ is  the 5D metric tensor $g_{MN}$, hence modifications of the bulk geometry produce variations in $T_{\mu \nu}$, which can be determined through a near-boundary expansion of $g_{MN}$
\cite{deHaro:2000xn}.

To implement  effects driving the boundary system out-of-equilibrium,   a distortion 
 (a {\it quench})  in the 4D metric can be introduced with profile $\gamma(\tau)$ \cite{Chesler:2008hg}, writing the boundary line element as
$
 ds^2=-d\tau^2+e^{\gamma(\tau)}dx_\perp^2+\tau^2 e^{-2 \gamma(\tau)}dy^2 \,\,.\label{metric4D}
$
 The corresponding 5D metric can be  expressed using Eddington-Finkelstein coordinates,  introducing  the fifth radial coordinate $r$ so that the boundary is reached for $r \to \infty$:
$
ds^2=2 dr d\tau-A d\tau^2+ \Sigma^2 e^B dx_\perp^2+ \Sigma^2 e^{-2B}dy^2 $.
 The  metric functions $A$, $\Sigma$, $B$  depend  on $r$ and $\tau$ only, due to  the imposed symmetries. 
 They can be computed   solving the Einstein equations with the   constraint that   the 4D metric  with  quench is recovered for $r \to \infty$.
Moreover, switching the quench on at $\tau=\tau_i$,   the metric functions must  reproduce  
 the  AdS$_5$ geometry at $\tau_i$.  A suitable expression for such equations has been worked out  in \cite{Chesler:2008hg}, and
an efficient solution algorithm  developed  in \cite{Bellantuono:2015hxa} has been applied to  different quench profiles $\gamma(\tau)$. 
The equilibration time  can be determined  comparing the behavior of the various observables  with the corresponding hydrodynamic quantities.  Here we describe the results for  two profiles,  denoted as model ${\cal A}_2$ and model ${\cal B}$,  which represent two different kinds of impulsive distortion of the boundary geometry. 

 \section{Role of local versus nonlocal observables to probe thermalization}\label{obs}
\begin{figure}[h!]
\begin{center}\vspace*{-.8cm}
\includegraphics[width = 0.43\textwidth]{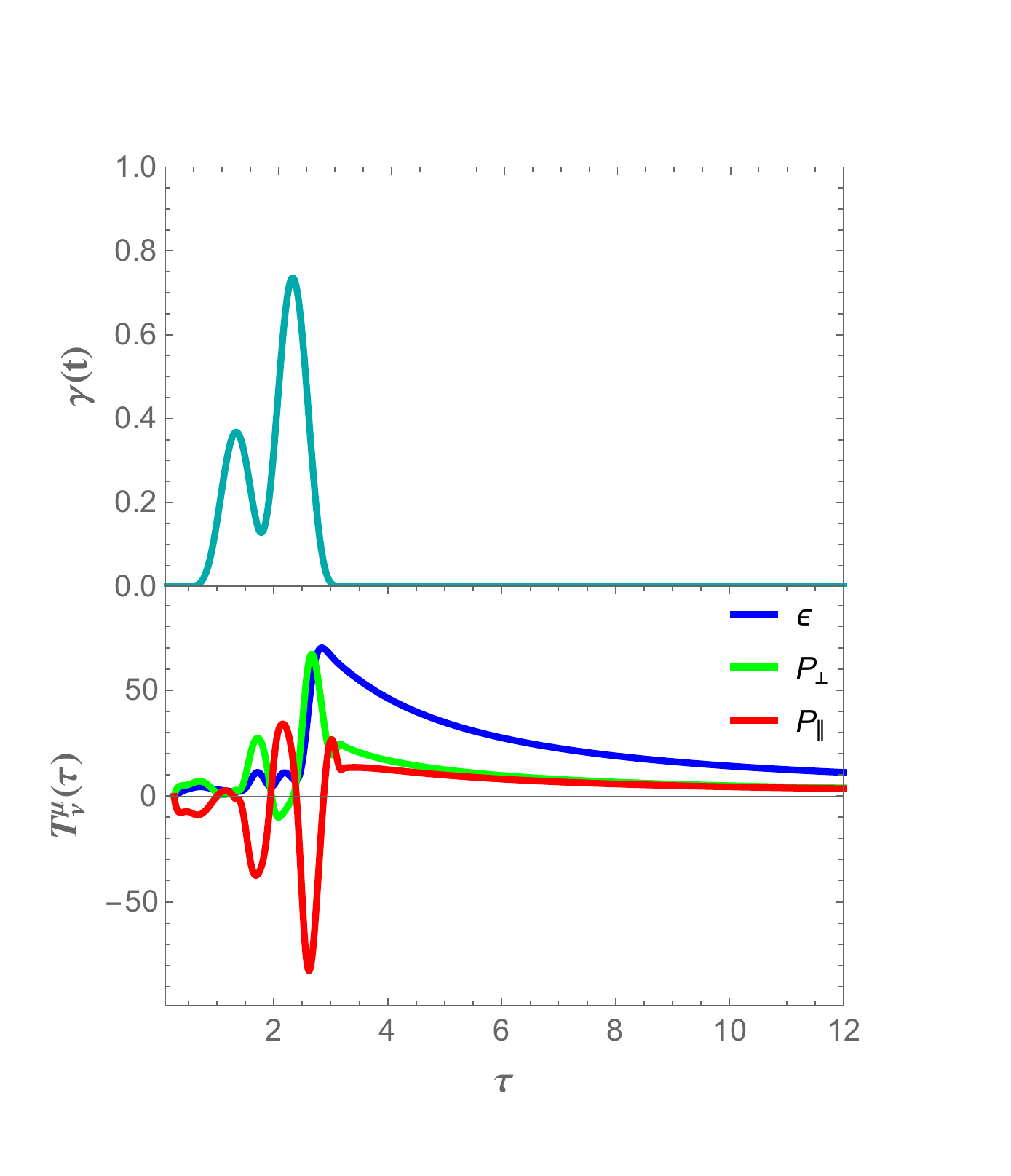}
\includegraphics[width = 0.43\textwidth]{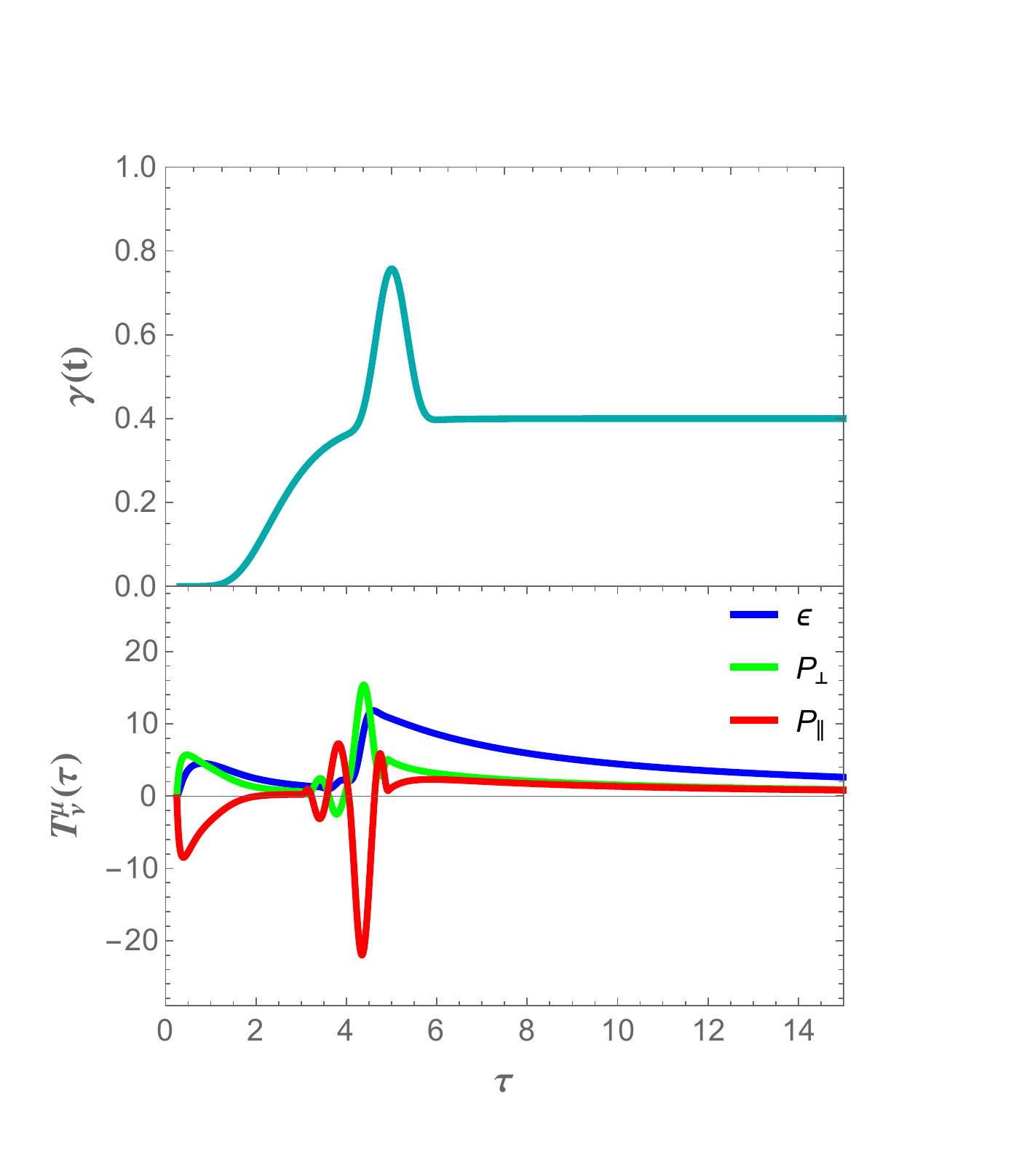}\\
\vspace*{-0.5cm}
\caption{\baselineskip 10pt  \small Profile $\gamma(\tau)$ (upper panel) and   components   of $T^\mu_\nu$  for the quench model ${\cal A}_2$ (left) and  ${\cal B}$ (right).  }\label{Tmunu}
\end{center}
\end{figure}

\begin{figure}[t]
\begin{center}
\includegraphics[width = 0.43\textwidth]{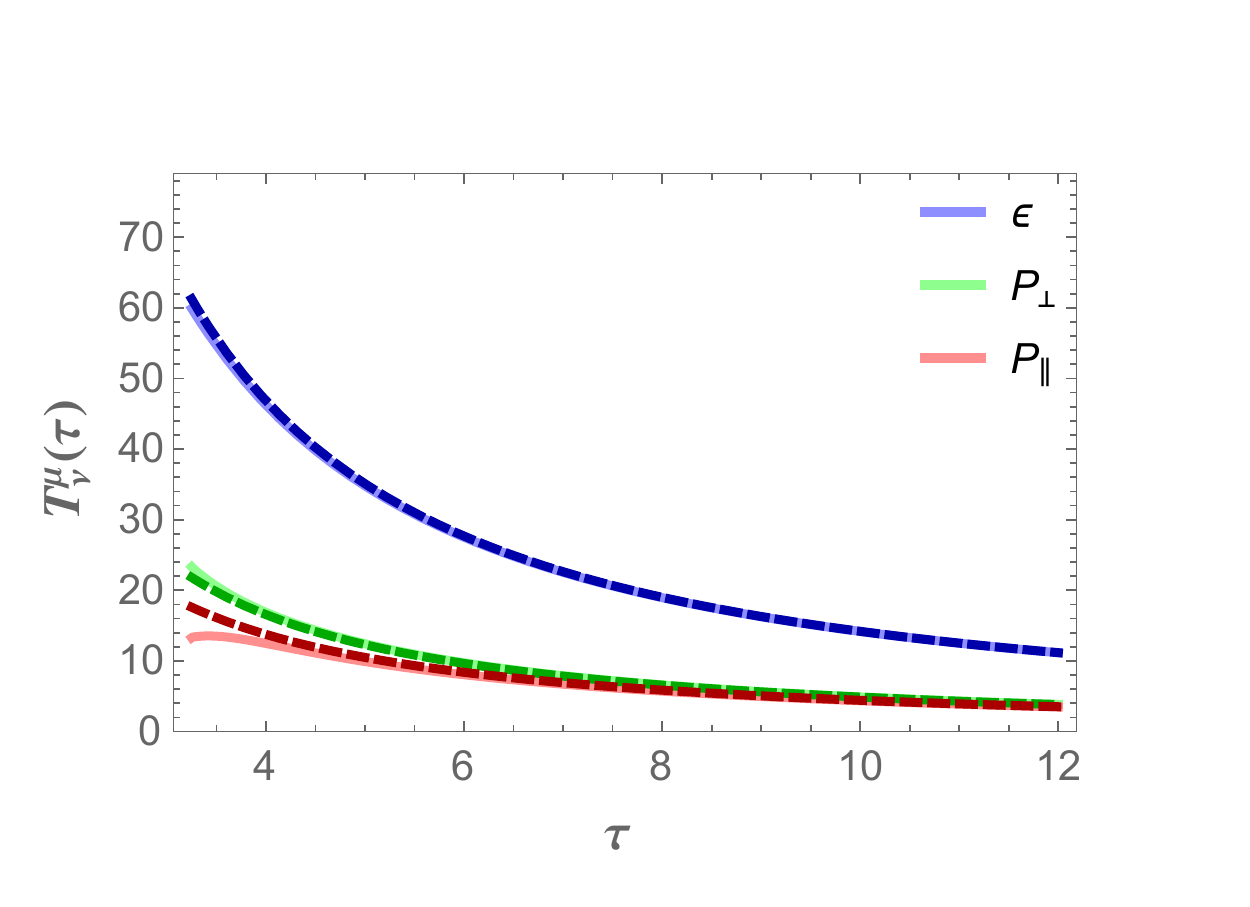}
\includegraphics[width = 0.43\textwidth]{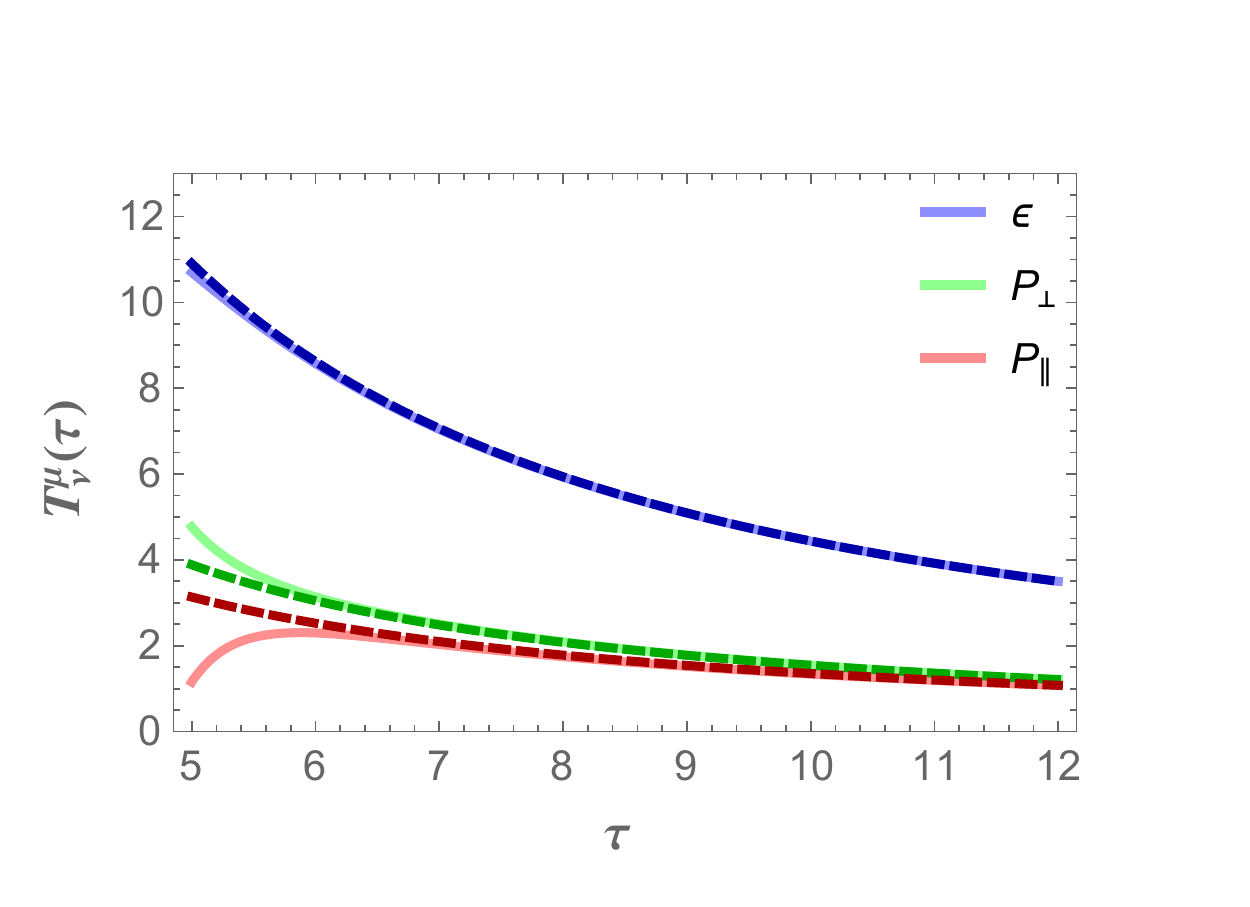}\\
\vspace*{-0.5cm}
\caption{\baselineskip 10pt  \small   $\epsilon(\tau)$, $p_\perp(\tau)$ and $p_\parallel(\tau)$ computed in model ${\cal A}_2$ for $\tau > \tau_f^{\cal A}$ (left) and in model ${\cal B}$ for $\tau > \tau_f^{\cal B}$ (right).  The  dashed lines correspond to the viscous  hydrodynamic expressions (1.1)-(1.3). }\label{Tmunuend}
\end{center}
\end{figure}

 The quench profiles for model  ${\cal A}_2$ and model ${\cal B}$ are depicted in the top panels of Fig.~\ref{Tmunu}. The deformation  persists up to $\tau_f^{\cal A}=3.25$ in model ${\cal A}_2$, and  up to $\tau_f^{\cal B}=5$ in model ${\cal B}$. 
 The lower panels  in the figure display the   $T_{\mu \nu}$ components \cite{Bellantuono:2015hxa}. 
 To investigate the late time  behavior,   Fig.~\ref{Tmunuend} shows the components of $T_{\mu \nu}$  after the end of the quench in comparison with  Eqs. (\ref{hydroeps})-(\ref{hydropperp}). 
 $\epsilon(\tau)$ follows the viscous hydrodynamics behavior right  after the end of the quenches, while a pressure anisotropy  persists  up to
$t_{isotr}^{{\cal A}}=6$ and  $t_{isotr}^{\cal B}=6.74$ \cite{Bellantuono:2015hxa}.
Imposing that at the end of the quench the temperature is
$T_{eff}=500$ MeV, the thermalization times turn out to be of ${\cal O}$(1 fm).
%
%

 The components of $T_{\mu\nu}$ are local observables mainly sensitive to the geometry close to the boundary. In  the holographic approach one can also  access nonlocal probes, namely  the two-point correlation function of boundary theory operators, and the expectation values of Wilson loops  on the boundary.
This requires the calculation of  the length of the geodesics in the bulk connecting the two boundary points in the correlation function, or the area of the extremal surface plunging in the bulk and having the Wilson loop as contour at the boundary  \cite{Balasubramanian:2010ce}.
  They have been  computed  for models ${\cal A}_2$ and  ${\cal B}$ and compared to
  the corresponding quantities in viscous hydrodynamics \cite{Bellantuono:2016tkh}. 
 The 5D metric reproducing the results  (\ref{hydroeps})-(\ref{hydropperp})
 has been worked out in   \cite{vanderSchee:2012qj,Bellantuono:2016tkh}.

\begin{figure}[b!]
\vspace*{-0.8cm}
\begin{center}
\begin{tabular}{ll}
\includegraphics[width = .43\textwidth]{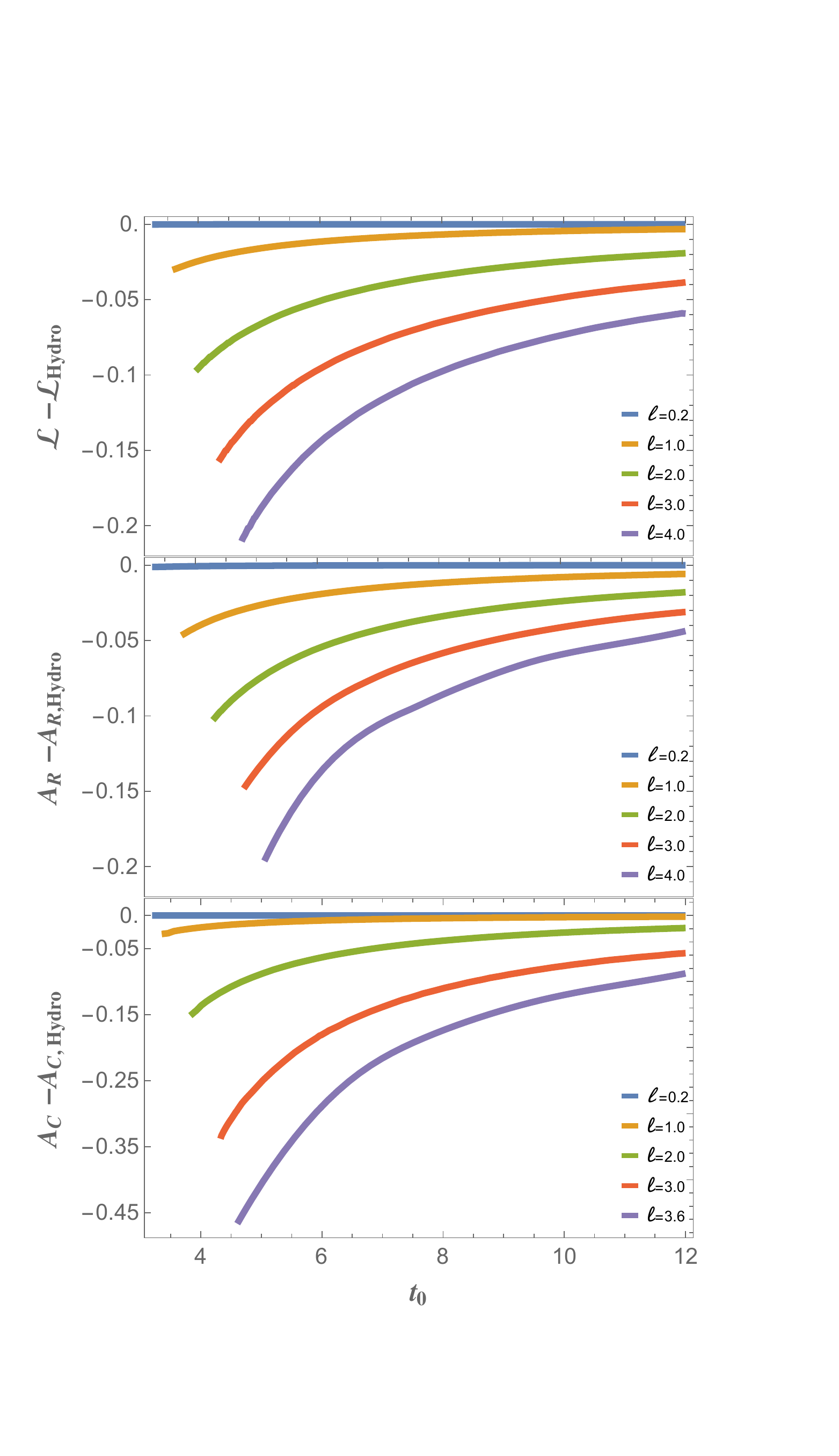}&
\includegraphics[width = .43\textwidth]{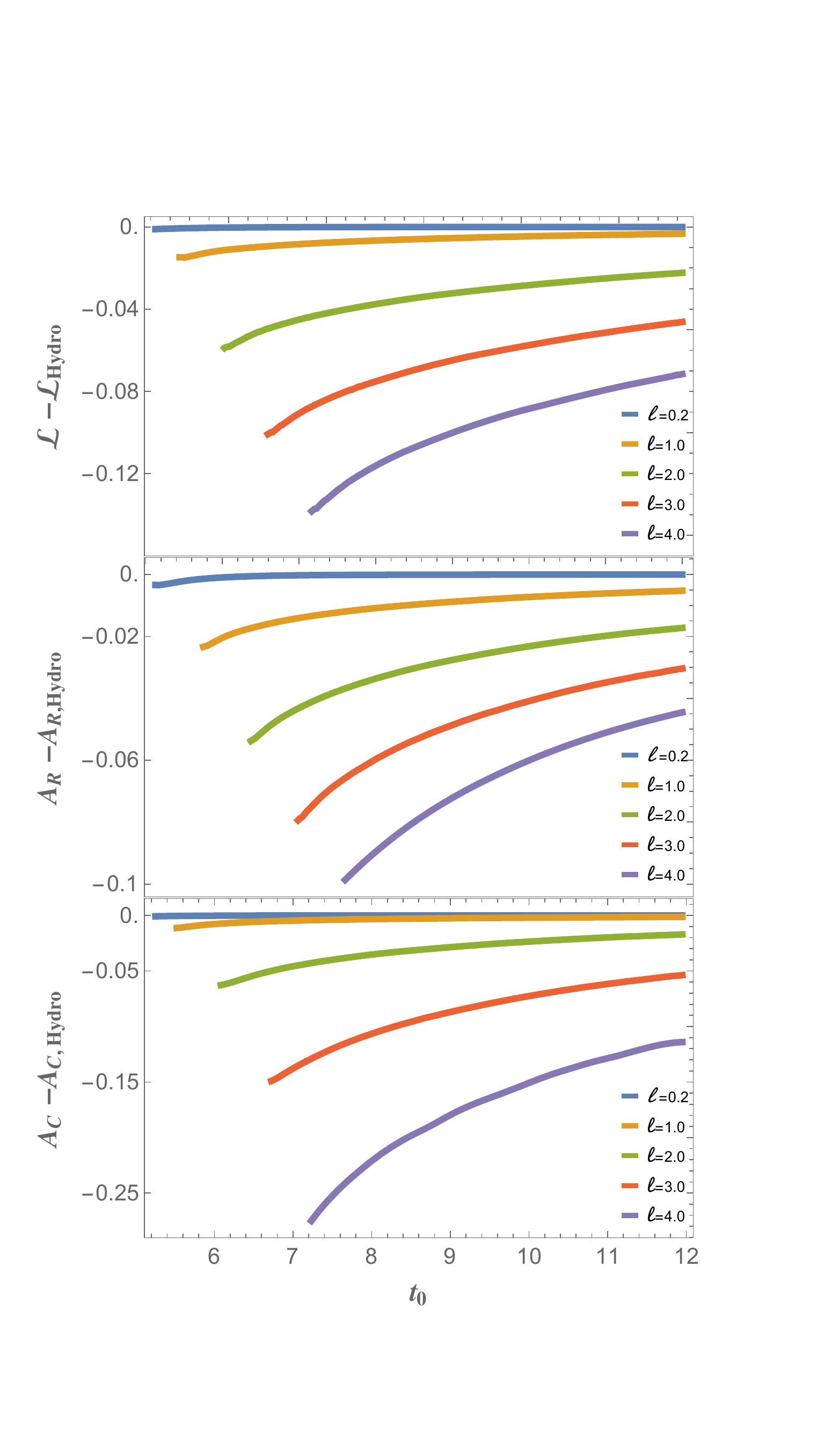} 
\end{tabular}
\vspace*{-0.5cm}
\caption{\baselineskip 10pt Results for model $\mathcal{A}_{(2)}$ (left) and $\mathcal{B}$ (right):   difference between the  geodesic length (top panel), the area of the extremal surface for the rectangular Wilson loop  (middle), and  for the circular Wilson loop (bottom)  in the models with quench and with the hydrodynamic metric. $t_0$ starts after the end of the quench. }\label{difnonlocal}
\end{center}
\end{figure}

\begin{figure}[t]
\begin{center}
\begin{tabular}{ll}
\includegraphics[width = .33\textwidth]{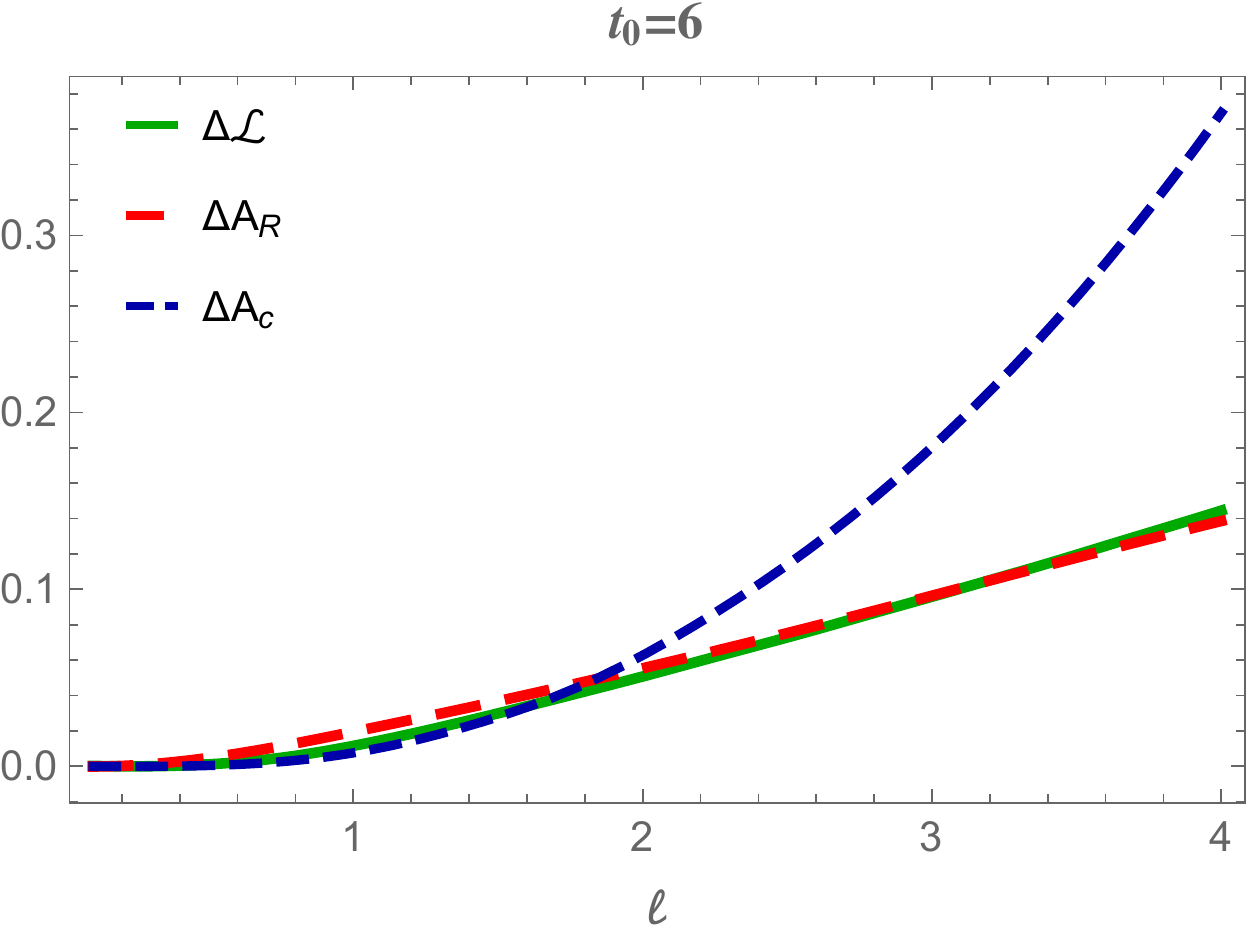} &
\includegraphics[width = .33\textwidth]{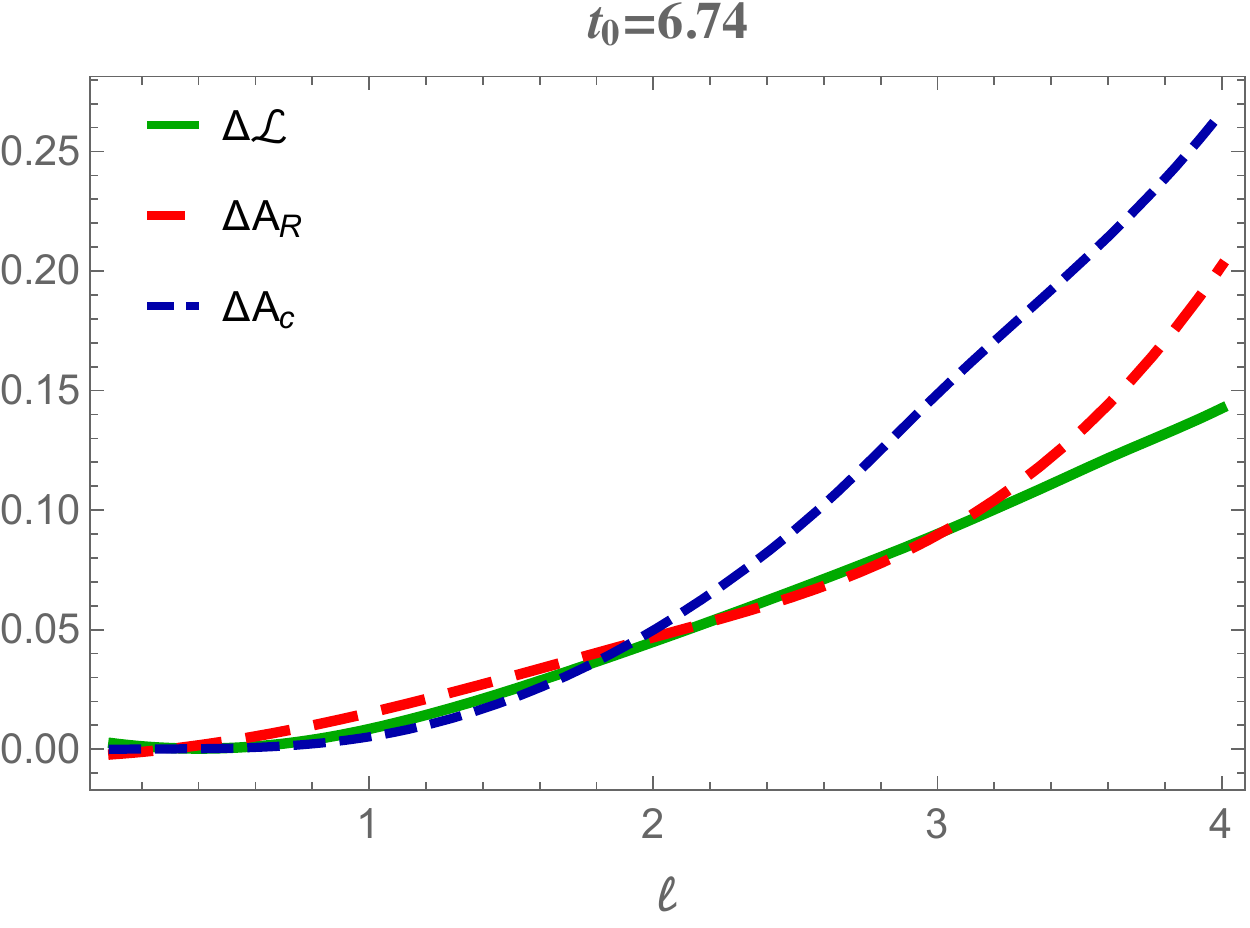} 
\end{tabular}
\vspace*{-0.2cm}
\caption{\baselineskip 10pt Differences $\Delta {\mathcal L}$, $\Delta A_R$, $\Delta A_C$   in model ${\cal A}_2$  at  $t_0=6$ (left),  and in model ${\cal B}$  at  $t_0=6.74$ (right) versus the size of the probe. }\label{674}
\end{center}
\end{figure}

The length of a curve connecting the  points $P$, $Q$ on the boundary is 
$\mathcal{L} = \int_{P}^Q d\lambda\sqrt{\pm g_{MN}\dot{x}^{M} \dot{x}^{N}}$,
where the coordinates $x^{M}(\lambda)$  depend on the parameter $\lambda$,     
and  $\dot{x}^{M}\equiv dx^{M}/d\lambda$ \cite{Balasubramanian:1999zv,Louko:2000tp}.
Viewing the integrand in $\mathcal{L}$ as a Lagrangian and solving the corresponding Euler-Lagrange equations, 
the   geodesic  can be   determined.
In the Eddington-Finkelstein coordinates,  the space-like path connecting the boundary points $P=\left(t_{0},-\ell/2,x_{2},y\right)$ and $Q=\left(t_{0},\ell/2,x_{2},y\right)$ that extends in the bulk at fixed $(x_{2},y)$  is described by the functions $\tau(x)$ and $r(x)$, with
$x_{1}\equiv x$, and
$\tau(0)=\tau_* $, $r(0)=r_* $, 
$\tau'(0)=r'(0)=0 $.
Boundary conditions are
$\tau(-\ell/2)=\tau(\ell/2)=t_{0}$, $r(-\ell/2)=r(\ell/2)=r_{0}
$.
In the calculation $r_0$ is set to  $r_0=12$.
The result for the geodesic length is
%
$\mathcal{L}=\displaystyle{\int_{-\ell/2}^{\ell/2} dx\frac{\tilde{\Sigma}(r,\tau)}{\sqrt{\tilde{\Sigma}(r_{*},\tau_{*})}} }$
%
with $\tilde{\Sigma}(r,\tau)\equiv\Sigma(r,\tau)^2e^{B(r,\tau)}$ 
computed in correspondence to
the solution $(r(x),\tau(x))$.
The calculation   for Wilson loops is described in \cite{Bellantuono:2016tkh}, where  two different shapes have been considered, a circle (C) and an infinite rectangular strip (R).
Fig.~\ref{difnonlocal}  displays for  models ${\cal A}_2$ and  ${\cal B}$ the differences $\Delta {\mathcal L}$, $\Delta A_R$ and  $\Delta A_C$ between each one of these three geometrical quantities  and  the corresponding quantity computed in the hydrodynamic setup,  after the end of the quenches. The thermalization time,  when the differences vanish, increases with the size of the probe, a feature of strongly coupled systems.
To compare  to the results from local probes,   fig. \ref{674}  shows $\Delta {\mathcal L}$, $\Delta A_R$, $\Delta A_C$ versus $\ell$ in model ${\cal A}_2$ at $t_0=t_{isotr}^{\cal A}=6$ and  in model ${\cal B}$ at $t_0=t_{isotr}^{\cal B}=6.74$: only for small   $\ell$ the differences are  zero.
%
 \section{Real-time quarkonium dissociation in the  far-from-equilibrium medium}\label{quarkonium}
\begin{figure}[t]
\begin{center}
\hspace*{0.5cm}
\includegraphics[width = 0.45\textwidth]{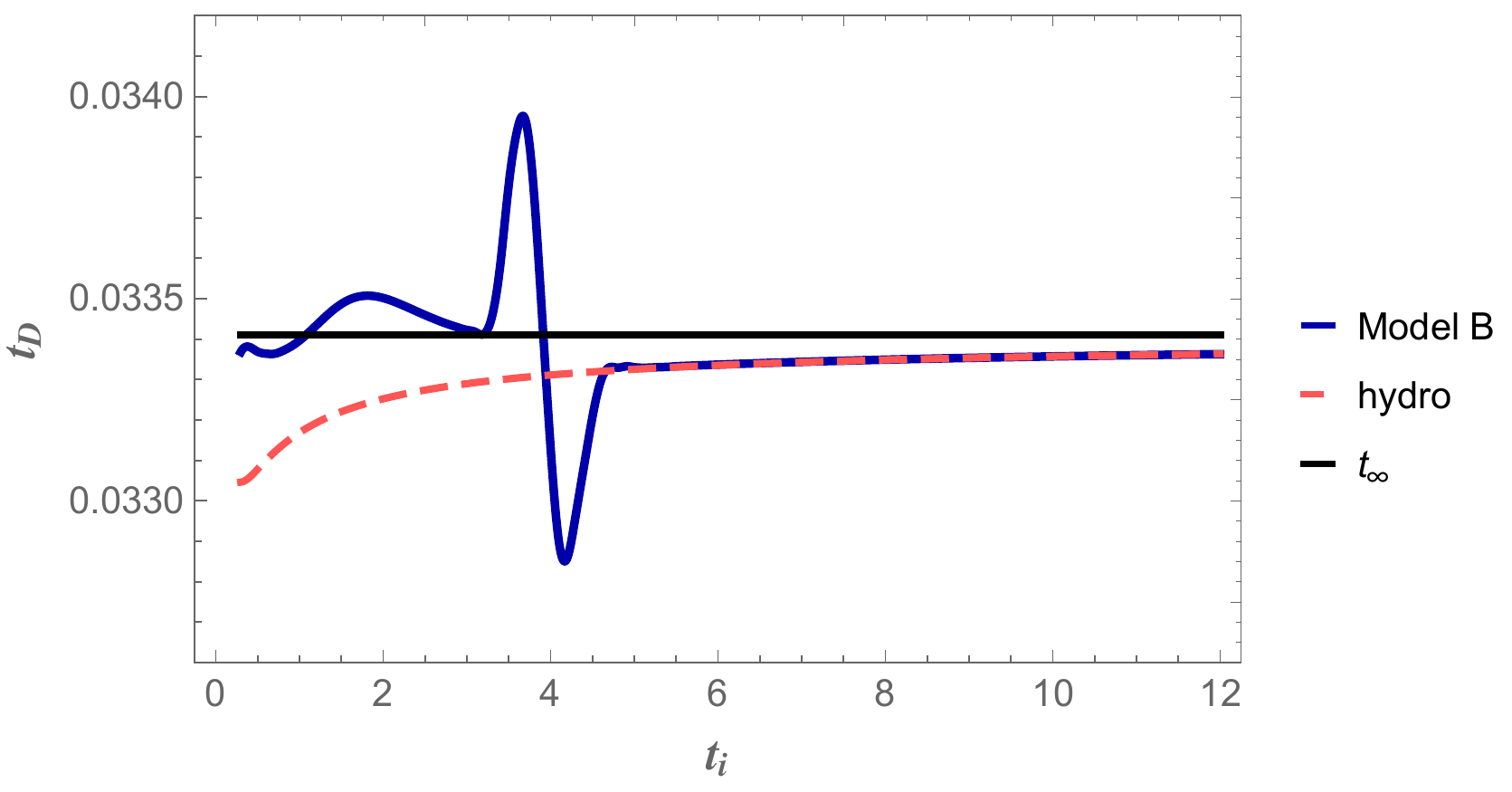}
\vspace*{-0.2cm}
\caption{ \baselineskip 10pt \small  $t_D$ versus  $t_i$ for    $w=x$, $L=10$ and $v=0$, for  quench model  ${\cal B}$ (continuous line) and 
 for a geometry dual to viscous hydrodynamics (dashed line).  The  horizontal line corresponds to  the asymptotic  value $t_\infty$.}\label{tDvsHydro}
\end{center}
\end{figure}

In the holographic approach quarks are dual to open strings in the bulk \cite{Karch:2002sh}.
In \cite{Bellantuono:2017msk} the real-time evolution of 
a string extending between two endpoints, representing a  heavy quark and an antiquark,  kept close to the boundary, is studied. 
The strings falls down under gravity and,   at finite temperature,
it can reach the black-hole horizon, an event   interpreted as the in-medium quarkonium  dissociation  \cite{Lin:2006rf,Iatrakis:2015sua}.
The   string dynamics is governed by  Nambu-Goto action $
S_{NG}=-T_f \int d\tau d\sigma \sqrt{-g} ,$
with $T_f=\displaystyle{\frac{1}{2\pi \alpha^\prime}}$, $\alpha^\prime=\frac{R_5^2}{\sqrt{\lambda}}$,  $R_5$ the AdS$_5$ radius and $\lambda$ the `t Hooft coupling. $g$ is the determinant of the induced world-sheet metric,   and $(\tau,\,\sigma)$  the world-sheet coordinates. 
We consider strings  in a 3D slice of the bulk described by the coordinates $(t,\,w,\,r)$,  with two choices for $w$.
 The first one is  $w=x$, with $x=x_1$ or $x=x_2$, and the string endpoints kept fixed at mutual distance $2 L$ close to the boundary. The second one is 
  $w=y$ along the rapidity axis, representing a quark and an antiquark  moving away  from each other  in the longitudinal direction $x_\parallel$ with rapidity $y_L$.
Choosing $\tau=t$ and $\sigma=w$, the string profile is a function $r(t,\,w)$.
In terms of the metric functions $A,\,B,\,\Sigma$, we find
$S_{NG}=-T_f \int dt \, dw \sqrt{\Sigma_w(t,r)\left(A(t,r)-2\, \partial_t r\right)+\left( \partial_w r \right)^2 }$,
where    $\Sigma_w={\bar \Sigma}=\Sigma^2 e^{-2B}$ if $w=y$ and $\Sigma_w={\tilde \Sigma}=\Sigma^2 e^{B}$ if $w=x$.
The resulting equation of motion for  $r(t,w)$ is ($r'=dr/dw$):
\be
r^{\prime \prime}-\frac{\partial_w g}{2g} r^\prime +\frac{\partial_t g}{2g}\Sigma_w-\partial_t \Sigma_w+\frac{\partial_r g}{2}=0 \,.
\label{eom} 
\ee
Since the  metric is 
 time-dependent, the solution depends on 
 the initial time  $t_i$ when the string is completely  stretched close to the boundary:
 $r(t_i,\,w)=r_{max}$ for all $w$ (we set $r_{max}=12$).
 The string endpoints  are kept  fixed at  $w_Q=-L$ and $w_{\overline Q}=L$ in the $w=x$ case, and  $w_Q=-y_L$ and $w_{\overline Q}=y_L$ for the $w=y$ configuration, so that
 $r(t,w_{\overline Q})=r(t,w_{Q})=r_{max}$.  We  vary $L$ and $y_L$ in the range $[0.1,\,100]$ and impose  the initial velocity $\dot r(t_i,w)=v$, with  $v=0,\,-0.5,\,-1$.
%
\begin{figure}[t!]
\begin{center}
\includegraphics[width = 0.47\textwidth]{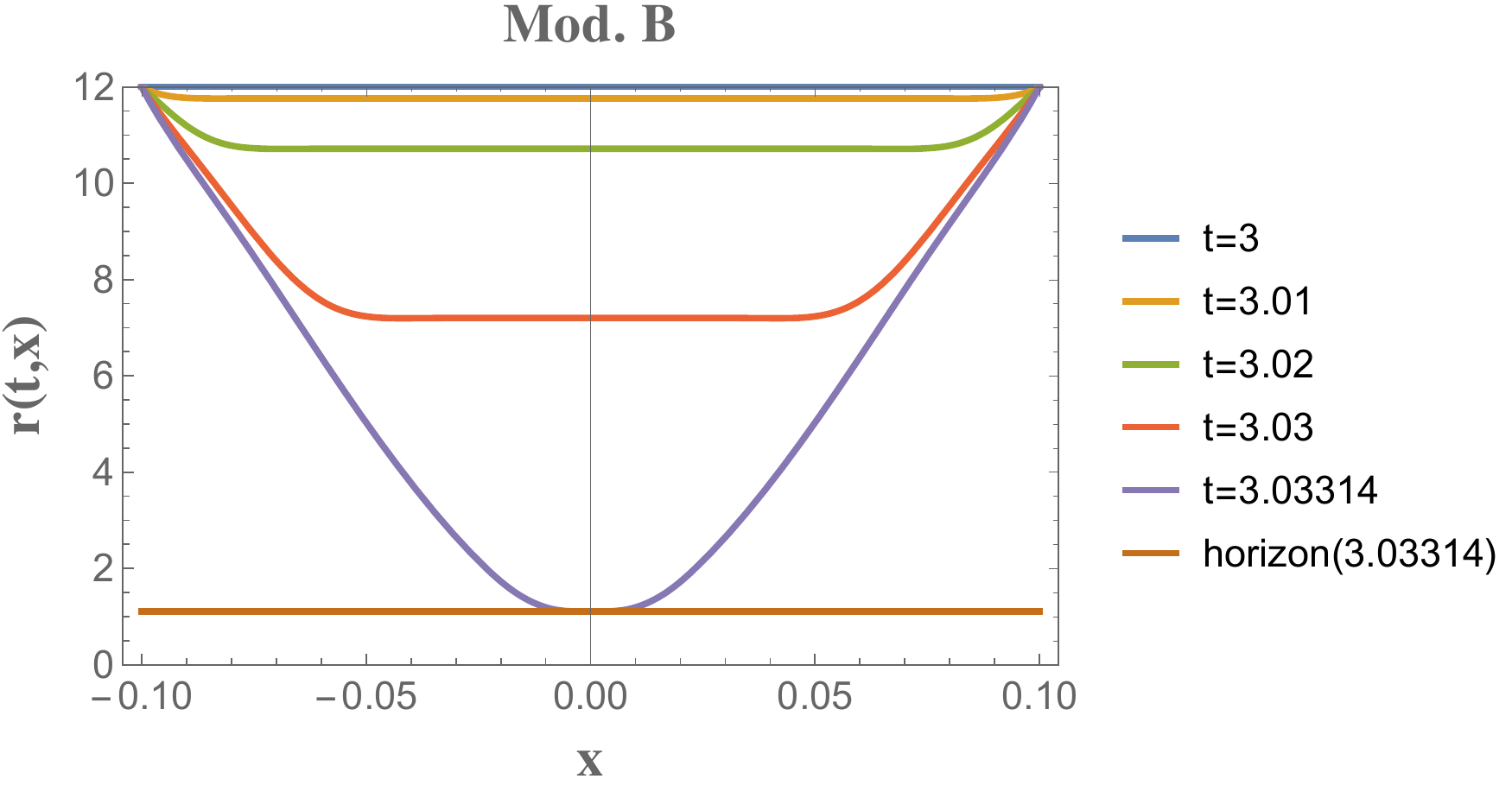}
\vspace*{-0.5cm}
\caption{\baselineskip 10pt \small 
String profile $r(t, w)$, corresponding to  $\{t_{i}, v, L\} = \{3, -1, 0.1\}$, for the transverse $w = x$ configuration and quench model $\cal B$ as a function of $x$ at different  $t$,  until the horizon is reached.  }\label{solutionsA2}
\end{center}
\end{figure}
%
%
\begin{figure}[b!]
\begin{center}\vspace*{-1.2cm}
\includegraphics[width = 0.43\textwidth]{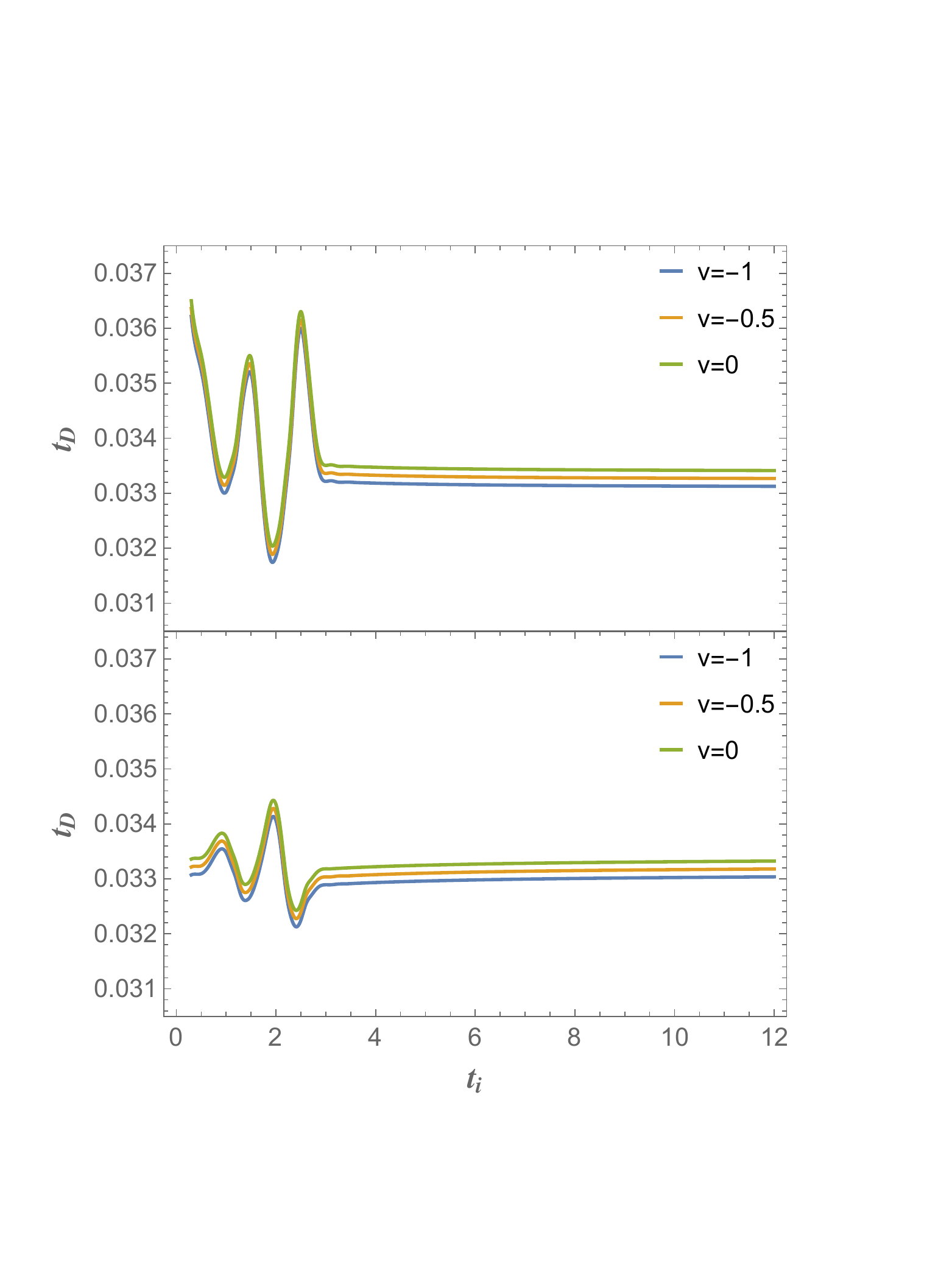}
\includegraphics[width = 0.43\textwidth]{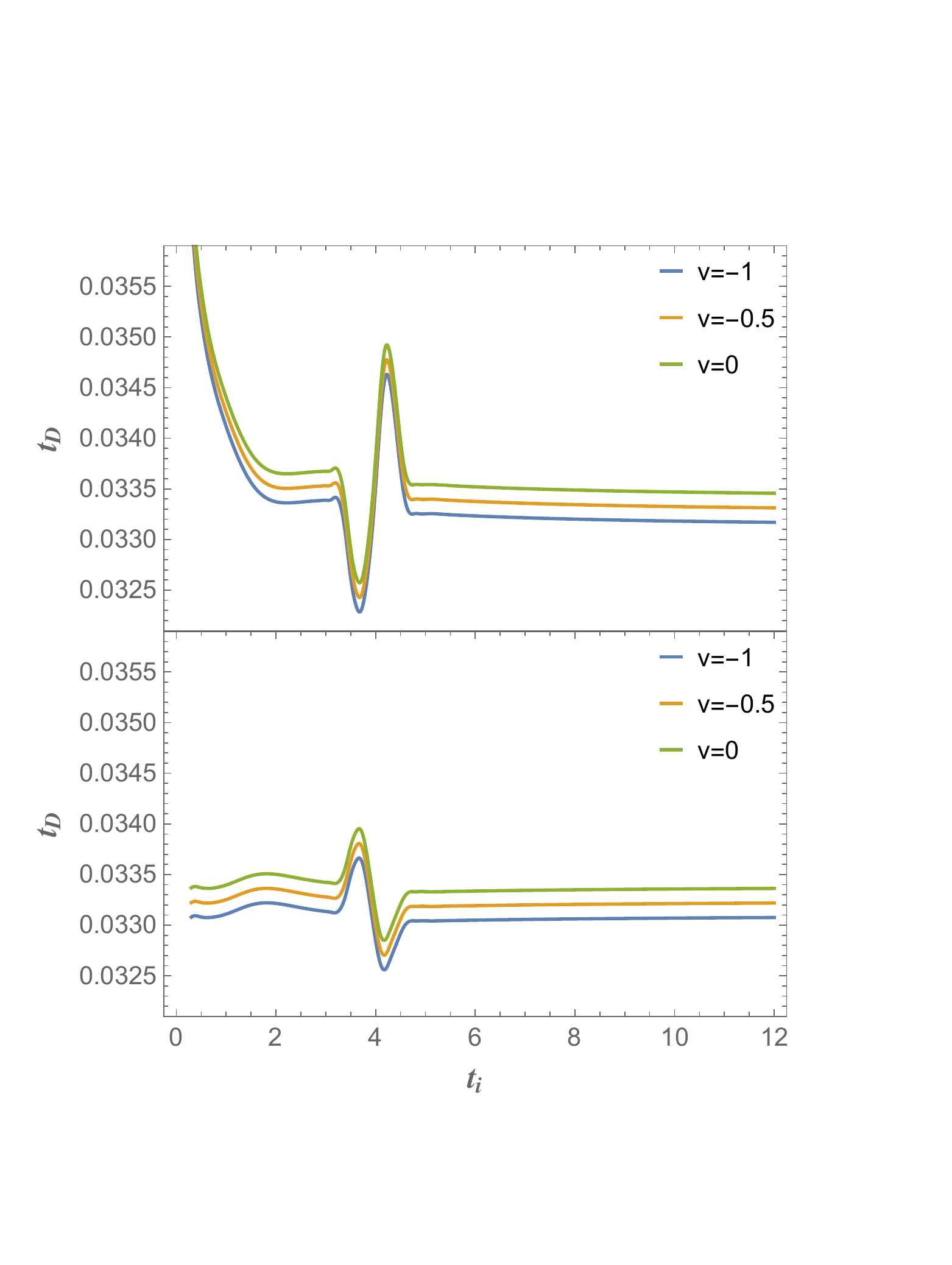}\\
\vspace*{-.5cm}
\caption{\baselineskip 10pt  \small  $t_D$ versus $t_i$ in  model ${\mathcal A}_{(2)}$ (left) and ${\cal B}$ (right). 
Top panels:    $w=y$  string configuration  with   $y_L=10$.    Bottom panels: $w=x$ configuration   with   $L=10$.   }\label{columns}
\end{center}
\end{figure}
%
Fig.~\ref{solutionsA2} displays  the string profile  in  model ${\cal B}$ for  $w=x$ and  $L=0.1$. Similar profiles are found for $w=y$ and in model ${\cal A}_{(2)}$. The dissociation time $t_D$ (finite in the chosen coordinate system) is determined when the string reaches the horizon.
 Fig.~\ref{columns} shows 
 $t_D$  versus  $t_i$ in  the two  models.
In each column, the upper panel displays $t_D$ for the  $w=y$ configuration  with  $y_L=10$, and the bottom panel  refers to $t_D$ in
the  $w=x$ configuration   with   $L=10$. 
After the end of the quenches, $t_D$ varies smoothly and approaches values close to each other in the two models. During the  quenches, $t_D$ abruptly  fluctuates
with a    different behavior for the two string configurations,  similarly to the  pressures   $p_\parallel$ and  $p_\perp$   \cite{Bellantuono:2015hxa} and to  the screening length
  \cite{Liu:2006nn}.
 Fig.~\ref{tDvsHydro}  compares the result for $t_D$ in  model ${\cal B}$ and in viscous hydrodynamics.
The hydrodynamic behavior   is  recovered right after the end of the quench,  and $t_D$ asympotically approaches the time required  to reach the AdS center starting from $r=r_{max}$:
 $t_\infty=\displaystyle{\frac{2}{3r_{max}} }\,_2F_1\left(1,\frac{5}{4},\frac{7}{4},-1\right)$.
\section{Conclusions}
Holographic methods allow us to describe 
the thermalization of a strongly interacting non-Abelian plasma,  driven out-of-equilibrium by a quench on the boundary geometry.   Local and nonlocal observables 
provide indications on the  thermalization time at various length scales. The  energy density follows the
hydrodynamic viscous behavior  after the end of the quench, while  the pressures take longer. For nonlocal observables  the thermalization
time  increases  with the size of the probe. 
Quarkonium dissociation is a fast phenomenon: the dissociation time follows the  behavior of viscous hydrodynamics as soon as the quench is switched off.

\vspace*{0.2cm}
These studies have been carried out within the INFN project QFT-HEP. LB thanks the  Angelo Della Riccia Foundation for financial support.

\end{document}